\documentclass[sigconf]{acmart}

\usepackage{pdfpages}
\usepackage{multirow}
\usepackage{enumitem}
\usepackage{tabularx,colortbl}
\usepackage{subcaption}
\usepackage{graphicx}

\AtBeginDocument{%
  \providecommand\BibTeX{{%
    \normalfont B\kern-0.5em{\scshape i\kern-0.25em b}\kern-0.8em\TeX}}}

\setcopyright{acmcopyright}
\copyrightyear{2023}
\acmYear{2023}
\acmDOI{XXXXXXX.XXXXXXX}

\acmConference[SIGIR 2023]{}{July 23--27,
  2023}{Taipei, Taiwan}
\acmPrice{15.00}
\acmISBN{978-1-4503-XXXX-X/18/06}

\begin{document}

\title[Detection of Messages Calling for Help in Earthquake Disaster]{Tweets Under the Rubble:\\Detection of Messages Calling for Help in Earthquake Disaster}


\author{Cagri Toraman}
\authornote{All authors contributed equally to this research.}
\email{ctoraman@aselsan.com.tr}
\orcid{}
\affiliation{%
  \institution{Aselsan Research Center}
  \streetaddress{Yenimahalle, 06400}
  \city{Ankara}
  \country{Turkey}
}

\author{Izzet Emre Kucukkaya}
\authornotemark[1]
\email{ekucukkaya@aselsan.com.tr}
\orcid{}
\affiliation{%
  \institution{Aselsan Research Center}
  \streetaddress{Yenimahalle, 06400}
  \city{Ankara}
  \country{Turkey}
}

\author{Oguzhan Ozcelik}
\authornotemark[1]
\email{ogozcelik@aselsan.com.tr}
\orcid{0000-0002-9420-9854}
\affiliation{%
  \institution{Aselsan Research Center}
  \streetaddress{Yenimahalle, 06400}
  \city{Ankara}
  \country{Turkey}
}

\author{Umitcan Sahin}
\authornotemark[1]
\email{ucsahin@aselsan.com.tr}
\orcid{}
\affiliation{%
  \institution{Aselsan Research Center}
  \streetaddress{Yenimahalle, 06400}
  \city{Ankara}
  \country{Turkey}
}

\renewcommand{\shortauthors}{Toraman et al.}

\begin{abstract}
The importance of social media is again exposed in the recent tragedy of the 2023 Turkey and Syria earthquake. Many victims who were trapped under the rubble called for help by posting messages in Twitter. We present an interactive tool to provide situational awareness for missing and trapped people, and disaster relief for rescue and donation efforts. The system (i) collects tweets, (ii) classifies the ones calling for help, (iii) extracts important entity tags, and (iv) visualizes them in an interactive map screen. Our initial experiments show that the performance in terms of the F1 score is up to 98.30 for tweet classification, and 84.32 for entity extraction. The demonstration, dataset, and other related files can be accessed at \href{https://github.com/avaapm/deprem}{https://github.com/avaapm/deprem}
\end{abstract}

\ccsdesc[500]{Computing methodologies~Natural language processing}
\ccsdesc[300]{Computing methodologies~Machine learning approaches}
\ccsdesc[300]{Information systems~Web and social media search}

\keywords{Earthquake, humanitarian aid, social media, tweet.}



\maketitle

\section{Introduction}
Social media is vital for situational awareness and humanitarian aids in the aftermath of natural disasters. The primary purpose is to save lives and reduce suffering. An example scenario was observed in the 2023 Turkey and Syria earthquake that affected a wide range including southern Turkey and northern Syria \cite{TheGuardian:2023a,BBC:2023}. After the earthquake, social media was used as a critical communication platform. Many people who were trapped under the rubble accessed to social media via mobile tools, and asked for rescue help by posting messages in Twitter \cite{TheGuardian:2023b,Euronews:2023,Northeastern:2023}. Many relatives and friends reported the missing people in the platform. And the need for different kinds of supplies and donations was raised in the following hours \cite{NYTimes:2023}.

\begin{figure}[t]
    \centering
    \includegraphics[clip, trim= 13cm 13cm 13cm 4cm, width=0.40\textwidth, page=1]{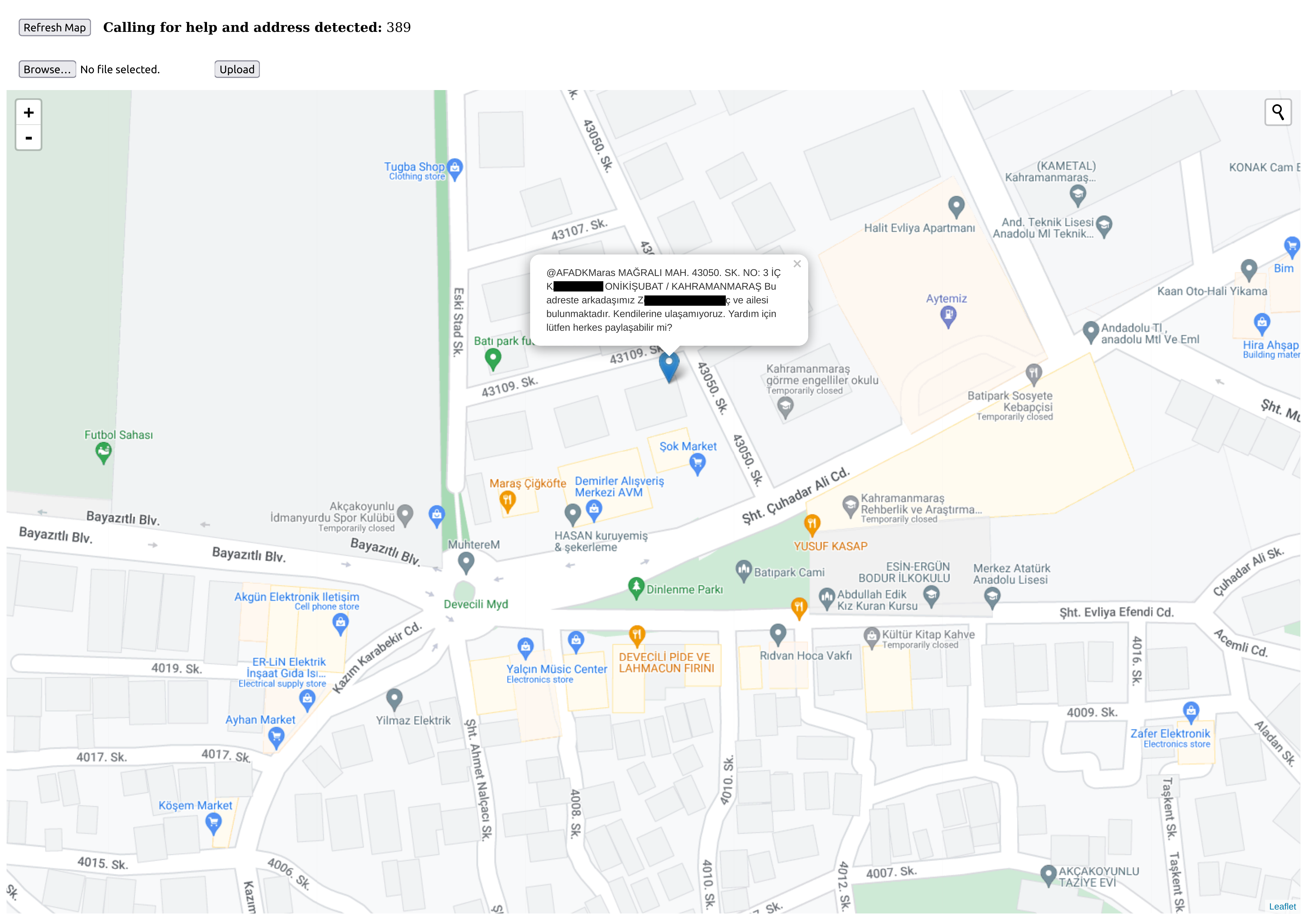}
    \caption{An example of located tweet calling for help. We remove any information that may reveal the personal identity in the figures throughout the paper to ensure privacy.}
    \label{fig:zoom-in-map}
\end{figure}

Humanitarian aids can be categorized under several classes \cite{imran2016lrec}. Our target classes in this study are "missing, trapped, or found people" and "rescue, donation efforts." In other words, we merge both classes under a coarse-grained representation, "calling for help" that represents social media messages seeking urgent help during the earthquake disaster (i.e. help for missing/trapped people and supply/donation). An example tweet calling for help is given in Figure \ref{fig:zoom-in-map}. 

\begin{figure}
    \centering
    \includegraphics[page=1, scale=0.55]{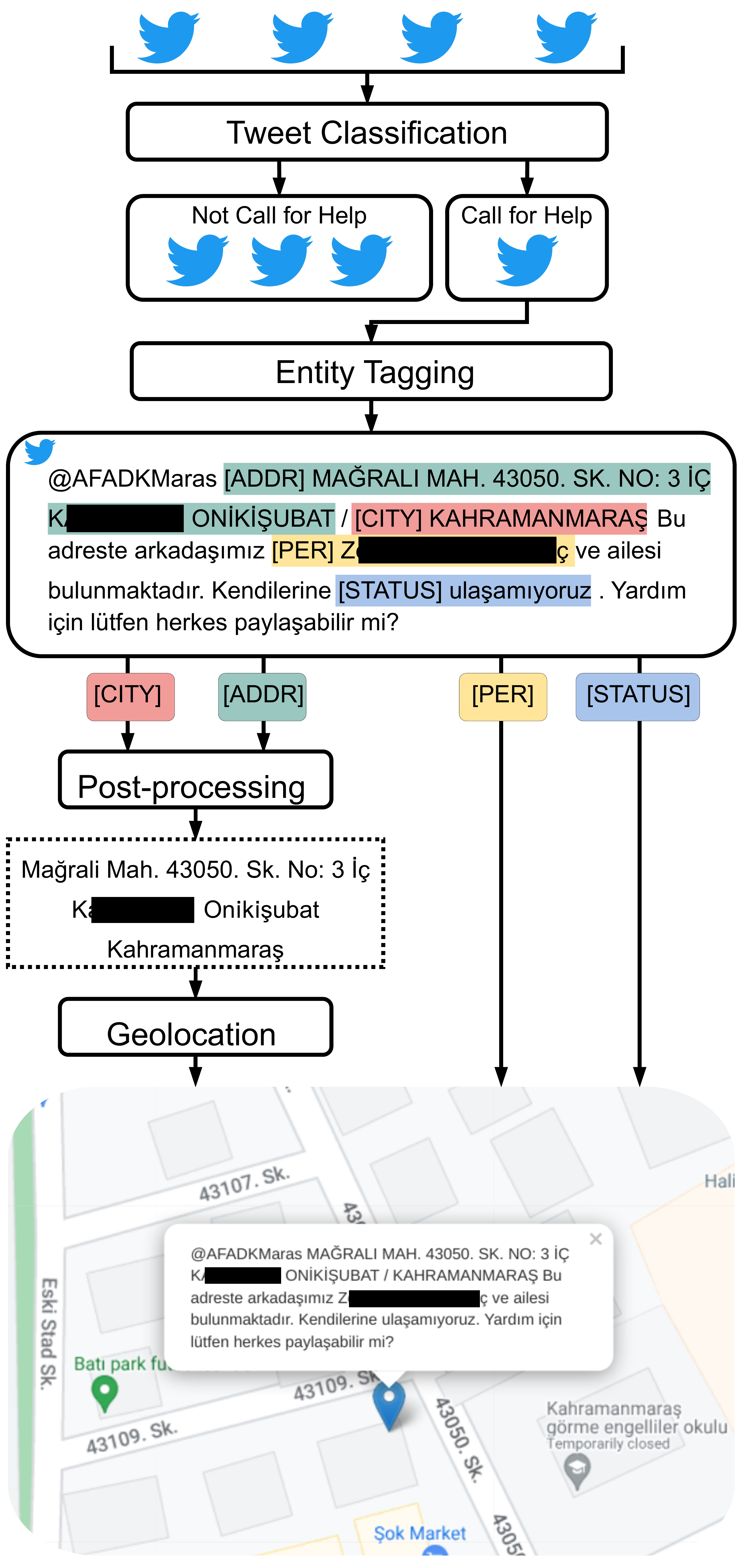}
    \caption{An illustration of the system pipeline for detecting tweets calling for help.}
    \label{fig:pipeline}
\end{figure}

We present an interactive tool that can provide situational awareness for missing and trapped people, and disaster relief for rescue and donation efforts. We are motivated by the tragedy of those affected in the 2023 Turkey-Syria earthquake. Indeed, we developed this tool in response to the earthquake in a few days. There are four consecutive tasks in the system. The first task is to collect tweets from Twitter API. We then have binary classification of social media messages (Turkish tweets in our case) whether they seek urgent help or not. Once tweets "calling for help" are determined, we extract important entity tags from their text. Specifically, we detect three entity tags: The person names, location (in terms of city and address), and status of the people who need help in tweets (e.g. under the rubble or missing). Lastly, we integrate the information that we extract from the previous steps into a graphical user interface, where tweets are placed on a map by using the location information (see Figure \ref{fig:pipeline}). We further provide the list of all predicted tweets along with the map for investigation purposes. Showing the tweets calling for help on a map along with a list of important tags would provide quick and accurate disaster response efforts.

\section{Related Work}
An existing dataset, HumAID, contains manually annotated tweets collected during 19 natural disaster events including earthquakes \cite{Alam:2021}. One of the class labels is "requests or urgent needs" that represents messages "of urgent needs or supplies such as food, water, clothing, money, medical supplies or blood." In this study, we extend this definition, and target the detection of social media messages of calling help, not only supplies but also requesting rescue operation (e.g. calling help under the rubble in an earthquake). Similarly, there are many datasets constructed for the analysis of various natural disasters and crises \cite{imran2016lrec, imran2013practical, imran2013extracting, firoj2018twitter, firoj_ACL_2018embaddings, ZAHRA2020102107, crisisdataset2020icwsm}. Furthermore, social media images are also widely used in analyzing crisis-related events \cite{nguyen2017image, crisismmd2018image, crisisdataset2020image}. 

There are many methods proposed to utilize the aforementioned datasets. Nguyen et al. \cite{nguyen16rapid} propose the use of convolutional neural networks (CNNs) for classifying crisis-related data on social media. Alam et al. \cite{crisisdataset2020image} create a benchmark for deep learning-based social media image classification in response to disasters. Zahra et al. \cite{ZAHRA2020102107} develop an automated approach for identifying eyewitness messages on Twitter during disasters using a combination of linguistic and contextual features. Moreover, Alam et al. \cite{firoj_ACL_2018embaddings} use domain adaptation with adversarial training and graph embeddings to classify social media posts after an earthquake. They use previous data from similar events (e.g. flood) to analyze unlabeled data for the current event. Different from the related studies, we utilize Transformer-based language models to classify tweets and extract relevant entity tags that request urgent assistance, and locate them on an interactive map for disaster relief. 

While developing our tool and writing this paper, we note that a group of independent collaborators together developed a web platform on which they extract information regarding people in need and collapsed building addresses\footnote{https://deprem.io/}, and share them on a visual map\footnote{https://afetharita.com/}, similar to the motivation of our efforts in response to the earthquake. We demonstrate our tool to the use of research community that can be further improved for the future disaster events. Moreover, we explain the development process in detail, and publish our dataset and related files for the community\footnote{The keywords and dataset are published at https://github.com/avaapm/deprem}.

\section{System Design}
The system is illustrated in Figure \ref{fig:pipeline} with the following consecutive tasks: Collecting tweets, tweet classification, entity tagging, post-processing, fetching geolocations, and lastly visualization in GUI. We explain each step in the following subsections.

\subsection{Data}

\subsubsection{Collecting Tweets}
In order to find relevant tweets, we used two sets of keywords\footnotemark[3]. The first set contains the general keywords that mentions about the earthquake, e.g. deprem (translated as "earthquake"). The second set contains the keywords calling for help, e.g. enkaz altinda (translated as "under the rubble"). We determined the keywords manually by browsing tweets in advance. We collected tweets via Twitter API's Academic Research Access\footnote{https://developer.twitter.com/en/products/twitter-api/academic-research}, ranging from the beginning of the earthquake to the end of the first 12 hours (i.e. from 04:00AM to 04:00PM on February 6th, 2023). The total number of collected tweets is approximately 1,824,000 (1,023,000 by querying help keywords and 801,000 by querying general keywords). We also collected 10,000 tweets by querying help keywords before the start of the earthquake to increase the complexity for training data.

\subsubsection{Annotation}
\label{subsec:annotation}
\begin{figure}[t]
    \centering
    \fbox{\includegraphics[clip, trim= 3cm 13cm 6cm 3cm, width=0.38\textwidth, page=1]{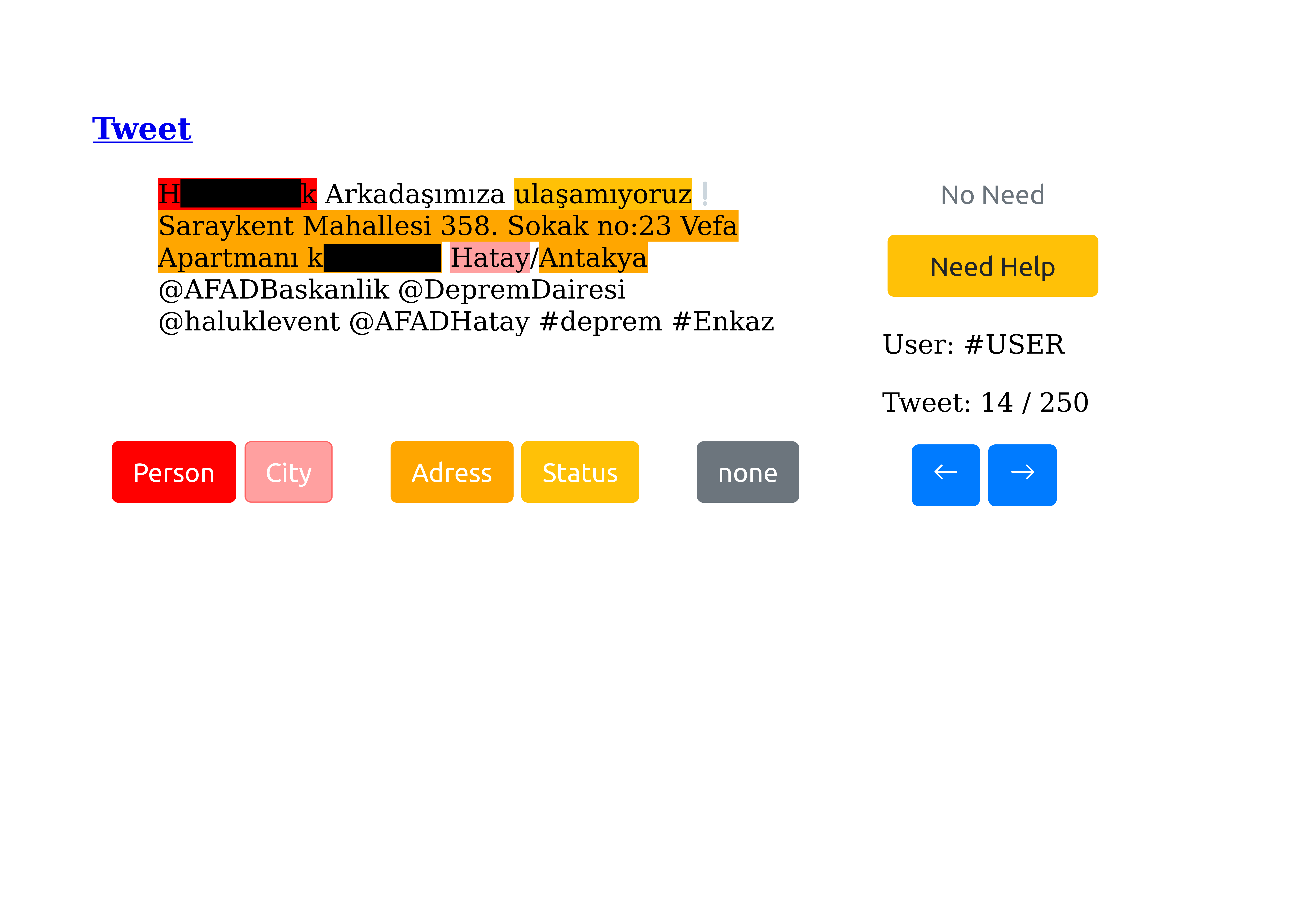}}
    \caption{A sample screen of the annotation tool.}
    \label{fig:annotation}
\end{figure}

We developed an annotation tool (a sample screen is given in Figure \ref{fig:annotation}). There are two different types of labels. The binary label determines if tweet calls for help (the buttons at the top right of the screen). The second type of labeling is entity tagging for the task of named entity recognition (the buttons at the bottom of the screen). The annotators highlight text segments and then select a type of named entity. There are four types of named entity. "Person" indicates the name of the person. "City" indicates the city of the help call. "Address" indicates the address of the help call. "Status" is the situation of the person who calls for help. "None" is used for clearing highlighted text. The annotation tool is based on Flask-Python\footnote{https://flask.palletsprojects.com} for the backend, and Javascript, CSS, and HTML for the frontend. The labels are kept in a database of sqlite\footnote{https://www.sqlite.org}.

We annotated 1,000 tweets for the training of natural language processing models. To provide data variety, we sampled 400 tweets from the general set, 400 from the help set, and 200 from the before-earthquake set. To develop the tool quickly in response to the earthquake, each tweet was labeled by a single annotator. There are four annotators who are the authors of this study, i.e. 250 tweets per annotator. The main statistics are given in Table \ref{tab:data_stats}.

\subsection{Natural Language Processing}

\subsubsection{Tweet Classification}
\label{subsec:training:classification}
We classify tweets that request urgent assistance for people that are trapped under the rubble, that are missing, or require emergency supplies and rescue. We compute TF-IDF vectors with a feature size of $7,703$ from the training data. Then, we train a Support Vector Machine (SVM) classifier using TF-IDF vectors. We also train multilingual and monolingual Transformer-based models for the binary classification, by choosing the models according to the observations in \cite{Sahin:2022}. As our monolingual model, we fine-tune a BERT model \cite{Devlin:2019}, called BERTurk\footnote{https://github.com/stefan-it/turkish-bert} pretrained on Turkish corpus. As our multilingual model, we fine-tune a mDeBERTa model, which is a multilingual version of DeBERTa, and uses the same structure as DeBERTa \cite{he2021deberta, he2021debertav3}. We use the \emph{base} and \emph{cased} versions of the monolingual and multilingual language models from the HuggingFace Transformers library \cite{Wolf:2020}. 

\subsubsection{Entity Tagging}
After classifying tweets that require urgent assistance, we extract predetermined coarse-grained entity tags to detect person name (\texttt{PER}), city (\texttt{CITY}), address (\texttt{ADDR}), and help status (\texttt{STATUS}) from tweet text. We implement Conditional Random Fields (CRF) \cite{Lafferty:2001} with hand-crafted features. For the hand-crafted feature set, we obtain (i) the stem of a word, (ii) Part-of-Speech tags of the current, previous, and next words, and (iii) Boolean indicators if the current words is titled, lower, or upper-cased. Furthermore, we utilize a Transformer-based monolingual model, ConvBERTurk\footnotemark[7] that is the Turkish pretrained version of ConvBERT \cite{Jiang:2020}, and a multilingual model, mDeBERTa \cite{he2021deberta, he2021debertav3}. The motivation behind choosing these models is based on the success of multilingual models in Turkish text and ConvBERTurk model in social media messages \cite{Ozcelik:2022}. We use the HuggingFace Transformers library \cite{Wolf:2020} for the model implementation. After obtaining entity tags, we post-process \texttt{ADDR} and \texttt{CITY} tags as follows. We create a list of cities affected by earthquake, and find the best match between the list in the first step and the extracted \texttt{CITY} tag using the Damerau–Levenshtein distance \cite{Damerau:1964}. The final address text is created with \texttt{ADDR} entities followed by \texttt{CITY} entity. We make address title cased, and remove non-alphanumeric characters except for dot, comma, and hyphen.

\begin{table}[t]
    \centering
    \renewcommand{\arraystretch}{1.2} 
    \begin{tabular}{l|r|rrrr}
         \multirow{2}{*}{\textbf{Binary label}} & \multirow{2}{*}{\textbf{Tweets}} & \multicolumn{4}{c}{\textbf{Entity classes}}\\
         & & \texttt{ADDR} & \texttt{CITY} & \texttt{PER} & \texttt{STATUS} \\ \hline \hline
         Call for Help & 418 & 516 & 504 & 364 & 392 \\ \hline
         Not Call for Help & 582 & \multicolumn{4}{c}{n/a} \\
         \hline
         \textbf{Total} & 1,000 & \multicolumn{4}{c}{1,776 entity tags} \\
         \hline 
    \end{tabular}
    \caption{The main statistics of the annotated data.}
    \label{tab:data_stats}
\end{table}

\subsection{Inference and GUI}
\begin{figure}[t]
\centering
\begin{subfigure}[b]{0.45\textwidth}
   \framebox{\includegraphics[clip, trim= 0cm 0cm 3cm 0cm,width=1\linewidth, page=1]{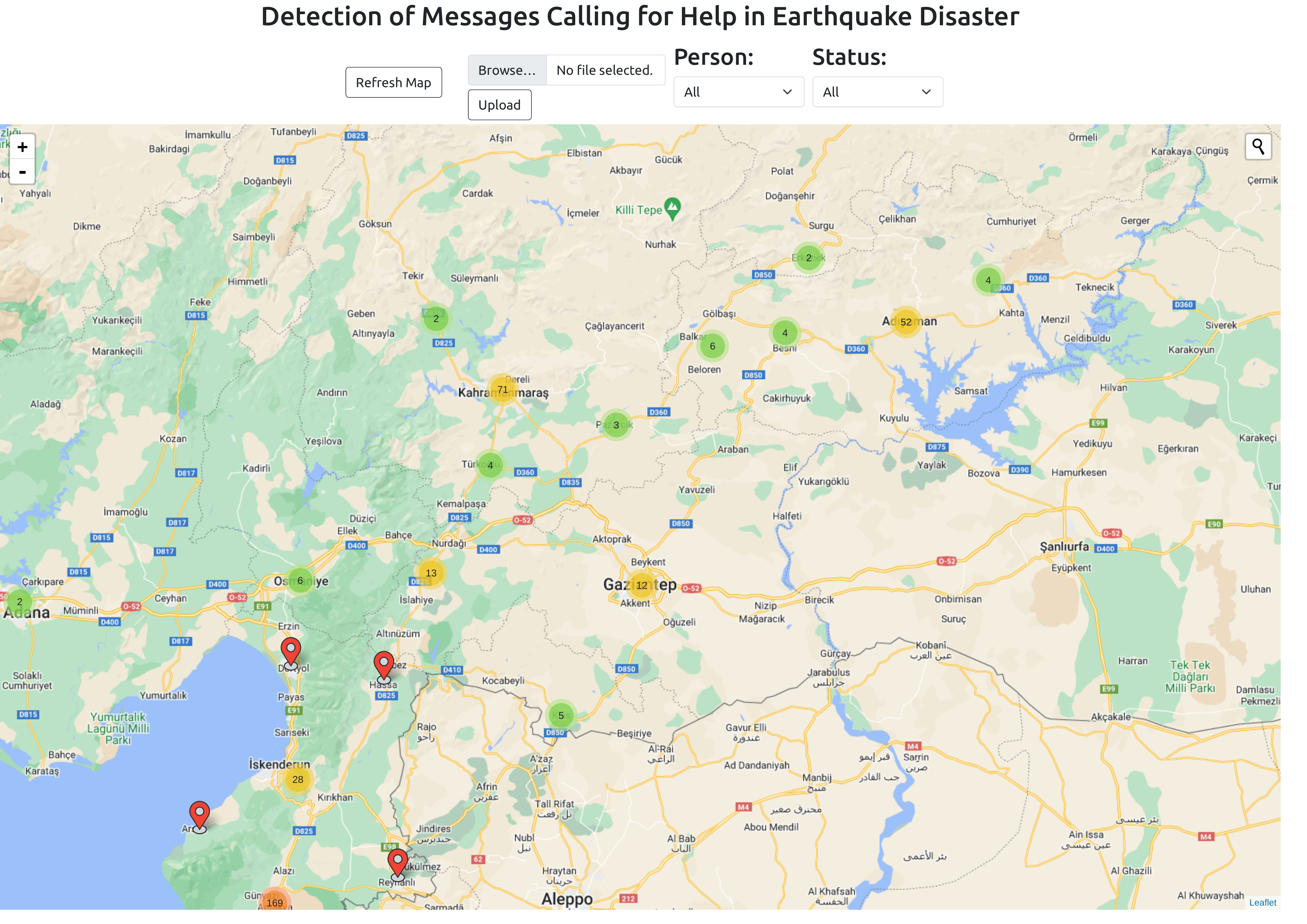}}
\end{subfigure}
\begin{subfigure}[b]{0.45\textwidth}
   \framebox{\includegraphics[width=1\linewidth, page=2, clip, trim= 0cm 12cm 1cm 0cm]{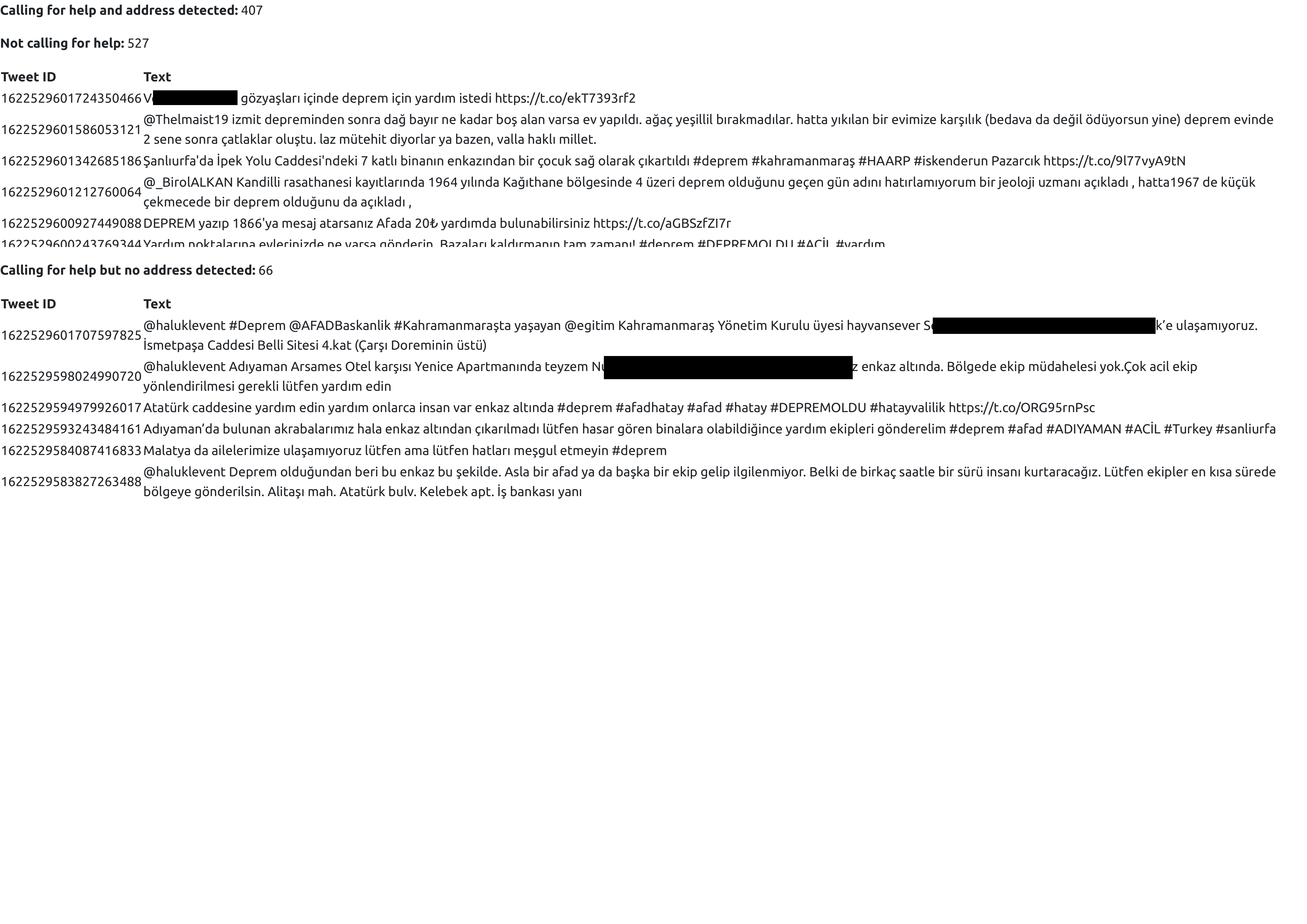}}
\end{subfigure}
\caption{A sample GUI screen where the tweets calling for help are marked (at the top), and tweets that are not located on the map for investigation purposes (at the bottom).}
\label{fig:gui}
\end{figure}

REST API\footnote{https://restfulapi.net} is used to process tweets from a post request as a JSON file containing tweet objects. We use the trained models to predict tweets calling for help along with entity tags. Using the address and city tags, the latitudes and longitudes of the locations are obtained from Google Maps\footnote{https://developers.google.com/maps/documentation/geocoding/}. Flask-Python is used to retrieve HTTP requests, and the fetch API is used for the GET requests to Google Maps API, which returns a JSON file containing the coordinate information. We filter the coordinates that are out-of-scope, i.e. wrongly located in a city that is out of the earthquake zone due to ambiguous addresses. We send the output to the graphical user interface (GUI). A sample GUI screen is given in Figure \ref{fig:gui}. The GUI is developed with Leaflet\footnote{https://leafletjs.com} and Google Maps. We provide two combo-boxes where user can select the names and statuses of people calling for help to see their location on the map. We also list all tweets, i.e. located and not located on the map for investigation purposes. 

\section{Evaluation}
In this section, we evaluate and report the model performance by using the annotated dataset. We choose the highest performing models\footnote{The models are published at https://huggingface.co/ctoraman} for deployment. For the demonstration, we use another set of unlabeled 1k tweets obtained from Twitter API in the same time range (first 12 hours after the earthquake).

\subsection{Tweet Classification}
For TF-IDF and SVM, we use the scikit-learn library with default parameters \cite{scikit-learn}. For BERTurk and mDeBERTa, we conduct hyper-parameter optimization by grid searching for the number of epochs in $[3, 5, 10, 20, 30]$, learning rates (LR) in [1e-5, 2e-5, 3e-5, 4e-5, 5e-5], batch sizes (BS) in $[2, 4, 8, 16, 32]$. During the fine-tuning, cross-entropy loss is used. For Transformer-based models, we use the tokenizers with \textit{longest} padding and \textit{truncation}. In terms of computational resources, the training was performed on three NVIDIA RTX 2080Ti. 

We apply stratified 5-fold splits on the training data (800 tweets for train, and 200 for test), and report the average F1 score of the positive class. We report the highest performing models with the best hyper-parameters in Table \ref{tab:tweet_class_results}. We observe that Transformer-based models perform better than SVM that is based on the bag-of-words model. The multilingual model, mDeBERTa, has competitive results to the monolingual model, BERTurk. We choose to deploy BERTurk for tweet classification, since it achieves the highest performance. 

\subsection{Entity Tagging}
We implement CRF with default hyper-parameters from scikit-learn \cite{scikit-learn} (i.e. during 1,000 iterations and 0.1 coefficient of L2 regularization for Stochastic Gradient Descent learning algorithm). Furthermore, we train ConvBERTurk and mDeBERTa with a fixed set of hyper-parameters by setting sequence length to 128 and learning rate to 5e-5 with a batch size of 8 during 10 epochs. Cross-entropy loss is used during the training. In terms of computational resources, the training was performed on a single NVIDIA RTX A4000. 

We apply 5-fold splits on the training data and report the average in terms of the weighted F1 score in Table \ref{tab:ner_results}. We use the standard CoNLL \cite{Connl:2003} metrics to measure the performances of named entity recognition models. The experimental results show that \texttt{ADDR} entity tag has lower scores compared to other entity tags in Table \ref{tab:ner_results}. The reason could be that the given addresses in tweets calling for help can be very long named entities (in terms of number of words), which can be difficult to predict without transition errors \cite{Ozcelik:2022}. The CRF model performs worse than the Transformer-based models. Interestingly, multilingual mDeBERTa model performs better than Turkish pretrained version of ConvBERT on average. This observation can show the generalization ability of multilingual models in specific languages compared to monolingual models. We choose to deploy mDeBERTa for entity tagging in our demonstration.

\subsection{Geolocation}
In the demonstration, we use 1k tweets as a simulation of tweet stream. 527 out of 1k tweets are classified by the trained model as "not calling for help." Among the remaining 473 tweets, our model can extract entity tags from 423 tweets. Lastly, 407 locations are fetched from the Google Maps API. Geolocations of 16 tweets could not be found (i.e. approximately 4\% is missed), possibly due to the ambiguity in the extracted address from tweets (e.g. missing the city name could result in locating an out-of-scope city having the same street name).

\begin{table}[t]
    \centering
    \begin{tabular}{l|rrr|r}
        \multirow{2}{*}{\textbf{Model}} & \multicolumn{3}{c|}{\textbf{Hyper-Param.}} & \multicolumn{1}{c}{\textbf{F1 Score}}  \ \\
        & EP & LR & BS & Avg. \\ 
        \hline \hline
        SVM & - & - & - & 89.40 \\
        BERTurk & 30 & 1e-5 & 8 & \textbf{98.30} \\ 
        mDeBERTa & 20 & 4e-5 & 2 & 98.05 \\ 
        \hline
    \end{tabular}
    \caption{Experimental results for tweet classification. The highest scores with best hyper-parameters are given in bold. \textit{EP}: number of training epochs, \textit{LR}: Learning rate, \textit{BS}: Batch size. We report the average of 5-fold splits in terms of the F1 score of the positive class.}
    \label{tab:tweet_class_results}
\end{table}

\begin{table}[t]
    \centering
    \small
    \setlength{\tabcolsep}{2pt}
    \begin{tabular}{l|rrr|rrrrr}
    \multirow{2}{*}{\textbf{Model}} & \multicolumn{3}{c|}{\textbf{Hyper-Param.}} & \multicolumn{5}{c}{\textbf{F1 Score}} \\
    & EP & LR & BS & ADDR & CITY & PER & STATUS & Avg.\\
    \hline \hline
    CRF & - & - & - & 55.63 & 88.65 & 58.79 & 76.67 & 70.67 \\ 
    ConvBERTurk & 10 & 5e-5 & 8 & 70.00 & 92.54 & \textbf{87.45} & \textbf{88.38} & 84.00 \\
    mDeBERTa & 10 & 5e-5 & 8 & \textbf{71.13} & \textbf{93.76} & 86.77 & 86.70 & \textbf{84.32}\\\hline
    \end{tabular}
    \caption{Experimental results for entity tagging. We report the average of 5-fold splits.}
    \label{tab:ner_results}
\end{table}

\section{Conclusion}
In response to the tragedy of the 2023 Turkey-Syria earthquake, we develop and demonstrate an interactive tool that can provide situational awareness for missing and trapped people, and disaster relief for rescue and donation efforts. The system consists of collecting tweets, classifying the ones calling for help, extracting entity tags, fetching geolocations, and lastly visualizing in a GUI. We belive that the system and demonstration can be further improved by integrating better performing models with enhanced training sets.

\begin{acks}
We dedicate this work to all people who were damaged in the 2023 Turkey and Syria earthquake.
\end{acks}



\end{document}